\documentclass[prd,aps,twocolumn,superscriptaddress,floatfix,tightenlines,nofootinbib]{revtex4}
\usepackage{amsmath}
\usepackage{amsfonts}
\usepackage{color}
\usepackage{units}
\usepackage{verbatim}
\usepackage{graphicx}
\usepackage{dcolumn}
\usepackage{amssymb}
\usepackage{bm}
\usepackage{marvosym}
\def\spose#1{\hbox to 0pt{#1\hss}}

\def\lta{\mathrel{\spose{\lower 3pt\hbox{$\mathchar"218$}}
\raise 2.0pt\hbox{$\mathchar"13C$}}}
\def\gta{\mathrel{\spose{\lower 3pt\hbox{$\mathchar"218$}}
\raise 2.0pt\hbox{$\mathchar"13E$}}}
\newcommand{\be}{\begin{equation}}
\newcommand{\en}{\end{equation}}
\newcommand{\bea}{\begin{eqnarray}}
\newcommand{\ena}{\end{eqnarray}}

\newcommand{\dd}{\mbox{d}}

\def\setR{\mathbb{R}}
\def\setC{\mathbb{C}}

\newcommand{\ie}{\textrm{i.e.~}}

\newcommand{\eg}{\textrm{e.g.~}}

\newcommand{\Ka}{\mathcal{K}}

\newcommand{\cs}{{c_{_\mathrm{S}}}}
\newcommand{\GN}{G_{_\mathrm{N}}}

\newcommand{\lP}{\ell_{_\mathrm{Pl}}}
\newcommand{\physrep}{Phys. Rep.}

\newcommand{\jcap}{JCAP}

\newcommand{\bl}{\color{black}}

\begin{document}

\title{Cosmology without inflation}

\author{Patrick Peter} \email{peter@iap.fr} \affiliation{${\cal
    G}\setR\varepsilon\setC{\cal O}$ -- Institut d'Astrophysique de
  Paris, UMR7095 CNRS, Universit\'e Pierre \& Marie Curie, 98 bis
  boulevard Arago, 75014 Paris, France}

\author{Nelson Pinto-Neto} \email{nelsonpn@cbpf.br} \affiliation{ICRA
  - Centro Brasileiro de Pesquisas F\'{\i}sicas -- CBPF, \\ rua Xavier
  Sigaud, 150, Urca, CEP22290-180, Rio de Janeiro, Brazil}

\begin{abstract}
  We propose a new cosmological paradigm in which our observed
  expanding phase is originated from an initially large contracting
  Universe that subsequently experienced a bounce.  This category of
  models, being geodesically complete, is non-singular and horizon-free,
  and can be made to prevent any relevant scale to ever have been
  smaller than the Planck length.  In this scenario, one can find new
  ways to solve the standard cosmological puzzles. One can also obtain
  scale invariant spectra for both scalar and tensor perturbations:
  this will be the case, for instance, if the contracting Universe is
  dust-dominated at the time at which large wavelength perturbations
  get larger than the curvature scale.  We present a particular
  example based on a dust fluid classically contracting model, where a
  bounce occurs due to quantum effects, in which these features are
  explicit.
\end{abstract}

\maketitle

\section{Introduction}

With the recent release of Wilkinson microwave anisotropie probe (WMAP)
data \cite{WMAP3,Komatsu+08}, the
inflation paradigm \cite{Staro79,MC812,Guth81,Linde82,inf25} has been
set on firmer ground.  Apart from solving some of the standard
cosmological puzzles (horizon, flatness, isotropy), the simplest
models predict an almost scale invariant spectrum of long wavelength
scalar perturbations, as observed, with low amplitude tensor
perturbations. This successful paradigm suffers, however, from some
weakening issues and omissions. The existence of an initial
singularity (a point where no physics is possible) in the standard
cosmological model is not addressed by inflation \cite{BV94}.  There
is no consensus yet as to whether inflation really solves the
homogeneity problem \cite{CS92,GP92} as long as one still needs
special initial conditions in a relatively large patch to initiate
inflation \cite{HW02,KLM02,GP92}. It seems that we cannot go forward
on this problem without a precise knowledge of how the Universe leaves
the Planck scale and/or a theory of initial conditions, \ie without
having an unambigous and complete theory of quantum gravity
\cite{GP92}. Furthermore, some cosmologically relevant wavelengths
must, at some early stage, have been trans-Planckian
\cite{MB01,BM01,ASPS01,LLMU02}; this can cast doubts on the validity
of the cosmological perturbation predictions of inflation.  Finally,
the usual and simpler models of inflation need a scalar field
\cite{brandenberger25}, whose theore\-tical properties demanded for
setting up the inflationary phase are not obviously compatible with
those obtained from well-motivated fundamental particle physics theory
\cite{lyth25,kallosh25} (a new perspective was, however, suggested
\cite{GMV08}).  In view of these difficulties, the question can be
asked whether the inflationary solution is unique.

Mechanisms that eliminate the initial singularity belong to one of the
following scenarios: either they assume a quantum creation of a small
but finite Universe and hence a beginning of time
\cite{HH83,Vilenkin86}, or they are based on an eternal Universe,
hence with no beginning of time. This last possibility can itself be
divided into two distinct categories: a monotonic time dependence of
the scale factor, \ie an expansion lasting forever, or different
phases including contractions and expansions, and therefore bounces.
The first situation is realised in the pre-big-bang (PBB) scenario
\cite{Veneziano91,GV03}; it requires a long accelerated phase
originating from either an asymptotically zero volume flat spacetime
or from a finite but small compact region \cite{MTLE05,FPS07} before
the usual decelerated expansion of the standard model. As for bouncing
models, they can be embedded in many theoretical situations
\cite{Murphy73,PF73,Novello08,NS79,MO79,dBPS98,CFP00,NAS08,BMS06,%
  ABFG02,BojowaldPRL01,BojowaldPRD01,ABL03}, including classically
singular cases \cite{KOST01,KKL01,MPPS03}.  In a string approach, both
situations are in practice equivalent due to the presence of the
dilaton which allows for a field reparametrization (as opposed to
conformal transformation as is usually, and erroneously, stated): the
PBB evolution of the Jordan (or string) frame is turned to a bounce in
the Einstein frame.

Up to now, there is not a single observation which favors one of these
three scenarios (time creation, eternal expansion or bounce) with
respect to the others, rendering these three possibilities susceptible
to physical investigation, without prior preferences.

Bouncing models differentiates, however, very strongly from the other
two scenarios above in one important aspect: initial conditions may
not be anymore put in a very small region, perhaps with Planckian
size, but in a very large and almost flat Universe. In this framework,
the flatness and the homogeneity problems are viewed from a very
different perspective.  Hence, bouncing models not only solve, by
construction, the singularity problem, but they may also possibly
solve, as we discuss in this work, other important puzzles of the
standard model without the need for an inflationary phase\footnote{Note
  that, strictly speaking, the bounce itself could be seen as
  including an inflationary phase since $\ddot{a}>0$ near the
  bounce. However, inflation is usually assumed (as we do in the
  present work) to be not only a period of acceleration, but one
  inducing many e-folds of expansion in a very short time. This is
  clearly not the case during a bounce.}.

Note also that a transition from contraction to expansion demands
non-standard or non classical physics, and/or non standard matter in order
to avoid the singularity in between. If, for having inflation,
violation of the strong energy condition is necessary and
sufficient, for bouncing models it may not be sufficient, requiring
also violation of the null energy condition in most cases (\ie
Friedmann models with nonpositive spatial hypersurface curvature
\cite{PP01}).  This suggests that there could possibly be
observational implications to which we shall come later. For now,
we turn to the way a bounce addresses the usual puzzles, before
presenting an actual model in which these features can be readily
implemented.

\section{Cosmological puzzles}

Bouncing models lead to a new framework for uncovering completely new
solutions to the standard cosmological puzzles. Let us list them in
what follows.

$\bullet$ Singularity: Bouncing universes are, by construction,
geodesically complete, and hence singularity free, so this point, not
addressed by inflation, is a non-issue here.

$\bullet$ Horizon: The size $d_{_{\mathrm H}}$ of the horizon is given
by the time integral $d_{_{\mathrm H}} (t) \equiv a(t)
\int_{t_\mathrm{i}}^{t} a^{-1}(\tau) \dd \tau$, with $t_\mathrm{i}$
some initial value. If the dynamics is driven by a perfect fluid with
constant equation of state $\omega\equiv p/\rho$, with $p$ and $\rho$
the pressure and energy density, respectively,
of the fluid, the scale factor
behaves, for flat hypersurfaces, as $a(t) \propto
|t|^{2/[3(1+\omega)]}$ [here and in what follows, we assume for
simplicity that the bounce takes place at $t=0$, so that $t<0$
($t>0$) represents the contracting (expanding) phase].

Integrating, we obtain the horizon as \be d_{_{\mathrm H}} =
\frac{3\left( 1+\omega\right) }{1+3\omega} \left\{
  \left|t_\mathrm{i}\right|^{(1+3\omega)/[3(1+\omega)]}
  -\left|t\right|^{2/[3(1+\omega)]} + t \right\}.
\label{horizon}
\en If $\omega>-\frac{1}{3}$, then clearly, as
$t_\mathrm{i}\to-\infty$ (bouncing case), $d_{_{\mathrm H}}$
diverges. At any finite time before or after the bounce, the horizon
is infinite and remains so for all subsequent times.  Note that this
solution would cease to be valid if, as seems to be the case now, the
Universe had been dominated by some kind of dark energy ($\omega <
-\frac{1}{3}$) in the contracting phase. This observation thus appears
to require a non symmetric bounce.

$\bullet$ Flatness: The problem stems from the classical equation
giving the density $\rho(t)$ relative to the critical one
$\rho_\mathrm{c} (t) = 3 H^2(t) /(8\pi\GN)$, with $H\equiv \dot a/a$
the Hubble expansion rate, through \be \frac{\dd}{\dd t} |\Omega-1| =
-2 \frac{\ddot{a}}{\dot a^{3}},
\label{Omega1}
\en where $\Omega \equiv\rho/\rho_\mathrm{c}$.  As $\Omega$ is close
to unity now, implying almost flat spatial sections (the term
involving the spatial curvature $\Ka$ would be negligible in the
Friedmann equation), Eq. (\ref{Omega1}) implies that it must have been
arbitrarily closer in the past in the usual big-bang scenario based on
decelerated ($\ddot a <0$) expansion ($\dot a >0$) since then
$|\Omega-1|$ is an ever-increasing function of time.  To solve this
problem, one must have had a long enough period during which
$|\Omega-1|$ decreases. This can be accomplished either through an
inflationary expansion phase ($\ddot{a}>0$ and $\dot{a}>0$) or
through a long decelerated contracting phase ($\ddot{a}<0$ and
$\dot{a}<0$). In the latter framework, we would say that the Universe
is seen to be almost flat now because it has expanded much less than
it has contracted before.

$\bullet$ Homogeneity: This is perhaps the deepest problem of the
standard model. There are essentially two approaches to this issue.
The first, exemplified here by the Weyl curvature hypothesis
\cite{Penrose79, PL00} (other examples on this approach have been
proposed \cite{HH83,Vilenkin86}, based on boundary conditions on the
wave function of the Universe), is to provide some theory of initial
conditions\footnote{Based on thermodynamical considerations, the Weyl
  curvature hypothesis consists in saying that the arrow of time
  implies the Universe to have an initially very low total entropy. It
  turns out that, if its gravitational part, the dominant one, depends
  only on the Weyl tensor, as the conjecture states, then it suffices
  to argue that the latter should be initially negligible. Note,
  however, that this particular hypothesis is not sufficient by itself
  to guarantee homogeneity as long as the conformal factor of the
  metric may have non-negligible spatial gradients at the
  beginning.}. The second possibility, of which inflation is
prototypical, is to invoke a dynamical process which wipes out any
preexisting inhomogeneity and anisotropy. In both cases, the outcome
should be the outstandingly special Friedmann-Lema\^{\i}tre-Robertson-Walker
(FLRW) geometry. It is
unquestionable that inflation, providing such a mechanism,
significantly alleviates the problem \cite{CS92,GP92,KLM02}, but it is
not clear whether it precludes special initial conditions
\cite{HW02,GP92} to be imposed.  It seems likely that a combination of
these two approaches will turn out to be necessary.  One expects that
whenever (if ever) a consistent theory of quantum gravity is
consensually accepted, it will, once applied to cosmology, provide the
required initial conditions to homogenize the primordial Universe.

In bouncing models, one may indeed envisage a solution for this
problem using a mixture of the above-mentioned two approaches within a
complete new perspective. In a very large and dilute Universe, the
energy-momentum tensor of matter, and hence the Ricci tensor, should
be very small. This requirement, by itself, does not ensure that the
geometry is almost flat since Einstein equations do not fix the Weyl
tensor. Under the Weyl curvature hypothesis, however, geometry should
be almost flat at that time, which we take to be our initial
condition. The question then becomes: do the initial inhomogeneities
grow?

Consider first the initial regime where the Universe is very large,
rarefied and, as discussed above, almost flat. Then the
self-gravitation of any inhomogeneity, even with $\delta \rho/\rho \gtrsim
1$, is negligible, as long as $\rho$ is very small. Note that this
would not be true if we were to take initial conditions at a time for
which the Universe is small and dense. These original inhomogeneities
therefore get dissipated in much the same way as air gets rapidly
homogeneous if perturbed (sound waves do not condense).  By
assumption, neither gravity nor its entropy are relevant in this
regime.  By the Hamilton theorem, the entropy of matter grows undisturbed
for a long time and thus can reach its maximum value.

Let us now go one step forward and assume a matter- (dust-)
dominated cosmological contraction.  In that case, the dust field velocity
evolves as $v\propto a^{-1}$, its number density $n\propto a^{-3}$,
and consequently its mean free path reads $\lambda_{_\mathrm{MFP}} =
(n\sigma)^{-1} \propto a^3$, where $\sigma$ is the dust cross section
(necessarily small for the dust approximation to make sense).

In a very large dust-dominated Universe, the Jeans length is
$\lambda_\mathrm{J}^\mathrm{ph}=\cs [\pi/(\GN\rho)]^{1/2}\propto
a^{1/2}$ and can be made larger than any large scale we see today. \bl
The dissipation time $t_\mathrm{d}$ for a given inhomogeneity of
wavelength $\lambda$ smaller than the Jeans length is given by \be
t_\mathrm{d}=\frac{\lambda}{v}\left(1+
\frac{\lambda}{\lambda_{_\mathrm{MFP}}}
\right),
\label{td}
\en and this time ought to be compared with the Hubble time scale. For
dust, $a\propto R_\mathrm{H}^{2/3}$, where we set $R_\mathrm{H} = y
R_0$ the Hubble radius at any time and $R_0$ its present value.
Writing $\lambda = x R_0$ and $\lambda_{_\mathrm{MFP}} = A
R_\mathrm{H}^2$, $A$ being a constant, Eq.~(\ref{td}) transforms into
\be t_\mathrm{d} \propto xy^{2/3} \left( 1+C \frac{x}{y^2}\right)
T_0, \label{dissip} \en where $T_0$ is the value of the Hubble time
today and $C^{-1} = A R_0$.  Comparing with the Hubble time
$t_\mathrm{H} = y T_0$, Eqs. (\ref{horizon}) and (\ref{dissip}) yield
\be \frac{t_\mathrm{d}}{t_\mathrm{H}} \propto \frac{x}{y^{1/3}} \left(
  1+C \frac{x}{y^2}\right).
\label{tdtH}
\en The dependence of (\ref{tdtH}) on $y$ obtained by the simple
calculation above shows that, for a sufficiently large $R_\mathrm{H} =
y R_0$, any scale up to the size of our Universe today becomes
homogeneous, being dissipated before gravity can play any role. In
fact, depending on the amount of time spent in this dust contraction
regime, and this time can be fixed arbitrarily large, even infinite if
one wishes, the dissipation is so effective that only quantum
fluctuations given by the uncertainty principle survive. This
provides, as a bonus, unique initial conditions for the perturbations:
vacuum fluctuations.

$\bullet$ Dark energy: This problem is not addressed by inflation, and
the simplest bouncing cases also remain silent here. However, as
discussed above, although dark energy is mostly harmless as inflation
proceeds, it may be problematic (see the horizon problem above) for
bouncing models. Hence, either dark energy was produced near or after
the bounce, or it cannot have dominated in the asymptotic past, as in
the transient dark energy example \cite{CALS06}.  In this case, one
could turn this potential difficulty into a means of reducing the
spectral index of perturbations: with a small amount of dark energy in
the primordial fluctuation enhancement epoch, the effective equation
of state could be made negative, thus implying a slightly red spectrum
(see below).  This is something to be investigated in more detail in
the future.

\section{Initial conditions for structure formation}

The main achievement of the inflation paradigm was the realization
that, due to the quantum fluctuations of the scalar field and the
metric, initial conditions for semiclassical perturbations could be
set in a natural way, demanding that at some stage the relevant scales
had been in a vacuum quantum state.  Implementing this condition then
led to the prediction that the spectral index of scalar perturbations
is close to 1 \cite{MC812}.  What similar initial conditions can be
imposed in bouncing models, if any, and what do they lead to in terms
of observations?

Setting vacuum initial conditions is in fact even more natural in a
bouncing case. Indeed, the Universe is supposed to be very large in
the far past and in fact, for the idea to make any sense at all, much
larger than any observable scale today. This means that, for any given
scale of interest, there exists a time, sufficiently before the
bounce, for which the scale in question is much smaller than the
curvature scale. As a result, one can safely work in the tangent
Minkowski space.  Furthermore, imposing vacuum for the corresponding
perturbations at that time is then not only a plausible requirement
but also a necessary consequence of the homogeneity solving scenario
discussed above, where inhomogeneities are dissipated up to quantum
vacuum fluctuations in a huge and slowly contracting Universe.

In the simple quantum cosmology background presented in the following
section, stemming from action (\ref{action}), explicit
calculations starting with vacuum initial conditions yields, for the
scalar and tensor spectral indices, respectively
\cite{ZN83,Wands99,PPP07,FB02}
\begin{equation}
\label{indexS} n_{_\mathrm{S}} = 1+\frac{12\omega}{1+3\omega},
\end{equation}
and
\begin{equation}
\label{indexT} n_{_\mathrm{T}} = \frac{12\omega}{1+3\omega}.
\end{equation}
In the dust (pressureless) limit $\omega\rightarrow 0$, one can easily
get a scale invariant spectrum for both tensor and scalar
perturbations, in agreement with observations. Furthermore, fitting
the amplitude of the perturbations with cosmic microwave background
(CMB) data leads to the nice
constraint that the curvature scale at the bounce should be greater
than roughly a thousand Planck lengths, ensuring that the model is not
spoiled by some discrete nature of spacetime such as induced by string
effects \cite{PPP07}.

The above calculations might erroneously lead one to believe that the
model necessarily involves only one fluid, and that it ought to be
dust at all times. Clearly, this would ruin the central idea.  In
fact, it is not mandatory that the fluid dominating the dynamics
during the bounce be dust. This is fortunate since densities and
temperature increase as the Universe contracts, eventually reaching
the point above which particle masses becomes negligible and the
Universe becomes radiation-dominated.  This would also happen as time
goes on if an initial bunch of monopoles and antimonopoles were to
annihilate. In any case, a matter to radiation transition is expected.

The reason why it is not necessary that dust dominate also during the
bounce is the following: the spectra of the growing and constant modes
of the Bardeen potential in the contraction phase are obtained far
from the bounce, and they do not change in a transition, say, from
matter to radiation domination (although amplitudes may change
\cite{MFB92}). The effect of the bounce is essentially to mix these
two coefficients (this also happens in other frameworks
\cite{FLP08,AP07}): the constant mode in the expansion phase is thus
very likely (although this is a model-dependent statement
\cite{Cai+07}) to acquire the scale invariant piece previously built
up. This happens whatever the fluid dominating at the bounce
\cite{PPP07}. Hence, the bounce may be dominated by any other fluid,
such as radiation. In short, providing the perturbations enter the
potential during an almost dustlike epoch, one expects the spectrum
to be almost scale invariant. We shall generally assume this
hypothesis, bearing in mind that it ought to be checked explicitly
afterwards \cite{MP042}.

Other bouncing representations have been discussed, among which
are purely
classical fluids, one of which, whose role is restricted to the bounce
itself, is of negative energy \cite{BV05,FPP}.  Such a negative energy
classical fluid might also be an effective fluid originated from
interactions among ordinary fluids in the early Universe
\cite{PF07}. Again, a scale invariant spectrum can be recovered
provided the positive energy fluid dominating at the early stage, when
the Universe is large, has an equation of state close to vanishing
(dust), irrespective of the negative energy fluid which drives the
bounce. Finally, and even though they are not mandatory, classical
scalar fields can also lead to bouncing models with a scale invariant
spectrum \cite{FB02}.  Hence this result is quite robust and not a
mere particular feature of a given specific model.

We now turn to our specific case which exemplifies all of the basic
requirements for the bounce paradigm we wish to defend as a would-be
``challenger'' to inflation. We would like to emphasize that, although
it possesses all of the features expected for a data-reproducing bounce,
its use merely serves the purpose of exhibiting how it can practically
be realized. As for inflation, many other solutions can be found, each
with its specificities; present \cite{WMAP3,Komatsu+08} or future
\cite{Planck} observational constraints might, however, hopefully
discriminate between the many possibilities.

\section{A quantum cosmological bounce}

What could be more simple for cosmology than to use Einstein action
sourced by a constant equation of state perfect fluid in 4 dimensions?
Amazingly enough, such an overwhelmingly simple framework manages to
reproduce all of the cosmic data, as we want to emphasize here. The
theory we deal with is thus \be {\mathcal S} = - \int \dd^4x \sqrt{-g}
\left( \frac{R}{6\lP^2} + \rho \right),
\label{action}
\en with $R$ the Ricci scalar and $\rho$ the energy density with
associated pressure $p = \omega \rho$, assuming $\omega$ to be a
constant.

We restrict our attention, to begin with, to homogeneous and isotropic
models and thus choose to consider the subset of metrics of the
FLRW form, namely, \be \dd
s^2 = g_{\mu\nu}^{(0)} \dd x^\mu \dd x^\nu = N^2(\tau) \dd \tau^2 -
a^2(\tau) \gamma_{ij} \dd x^i \dd x^j,
\label{FLRW}
\en with $\gamma_{ij}= \left( 1+\frac14\Ka \mbox{\boldmath $x$}^2
\right)^{-2} \delta_{ij}$ the 3-space metric and $a(\tau)$ the scale
factor. Note that we do not assume flat spatial sections, so the
spatial curvature $\Ka$ is free, although normalizable: $\Ka\in\left\{
  0,\pm1\right\}$. Finally, the lapse function can be chosen as
$N=a^{3\omega}$, so that $\tau$ is identified with cosmic time if the
fluid is made of dust, and conformal time if it is made of radiation.

Going on to perturbations around such a background, we write the full
metric as \be \dd s^2 = \left( g_{\mu\nu}^{(0)} + \delta
  g_{\mu\nu}\right) \dd x^\mu \dd x^\nu.
\label{pert1}
\en This, in principle, provides a full set of quantum
observables. Let us consider an arbitrary quantum state $\psi\left[
  g_{\mu\nu}\left( x\right), \cdots \right]$, where the dots stand for
whatever other degree of freedom is present. Consistency of the linear
quantum perturbation approach in this case might be asserted, or at
least addressed, provided that, for non vanishing values of the
background expectation values, the constraint \be \langle \psi |
\delta g_{\mu\nu} |\psi \rangle \equiv \langle \delta g_{\mu\nu}
\rangle_\psi \ll \langle g^{(0)}_{\mu\nu} \rangle_\psi
\label{consis}
\en
holds.

In the perturbed FLRW case in the longitudinal gauge, considering
scalar and tensor perturbations only\footnote{Both at classical and
  quantum levels, scalar, vector and tensor perturbations decouple,
  and vectors are rapidly diluted away by the expansion as $\propto
  a^{-2}$; they are not measurable today, so one merely needs to ensure
  they never spoiled the linear approximation.}  \be \dd s^2 = N^2
\left(1+2\Psi \right) \dd \tau^2 - a^2 \left[ \left(1-2\Phi \right)
  \gamma_{ij} + h_{ij} \right]\dd x^i \dd x^j,
\label{gauge}
\en the constraint (\ref{consis}) amounts to $\left( \langle \Psi
  \rangle_\psi , \langle \Phi \rangle_\psi , \langle h_{ij}
  \rangle_\psi \right) \ll 1$. Note that the tensor modes are
traceless and divergence-free, \ie $\gamma^{ij} h_{ij} = 0$ and
$h^{ij}_{\ \ ;j} = 0$, with the covariant derivative taken with
respect to $\gamma_{ij}$.

The crucial point concerning this expansion is that it can be shown
\cite{PP07} that the Fourier modes of these perturbations, in the
restricted case of a constant equation of state perfect fluid, satisfy
equations of motion that are exactly those of the classical theory.
In fact, at least in the flat $\Ka=0$ situation, they can be obtained
without any appeal to the background field equations and therefore
can be used straightforwardly in the quantum regime
\cite{PP07,PPP05,PPP06,PPP07}.  A consistent Hamiltonian constraint
$\mathcal{H} = \mathcal{H}_0 + \delta \mathcal{H}$ was obtained, where
$\mathcal{H}_0$ describes the background geometry while
$\delta\mathcal{H}$ is the Hamiltonian constraint for the
perturbations written in very simple form and suitable for Dirac
quantization.

The way to proceed is to go a step forward with respect to the usual
approach, where perturbations are quantized and the background remains
classical, and use the whole Hamiltonian constraint above to Dirac
quantize both the background and the perturbations\footnote{An attempt
  in this direction was done \cite{HH85}, which, however, could not be
  taken much forward due to the complicated form of $\delta\mathcal{H}$
  they use.}, making a wavefunction separation into zeroth and second
orders as \be \psi = \psi^{(0)} \left( a,\tau\right) \times \psi^{(2)}
\left[ a, \Psi \left( \mbox{\boldmath $x$}\right), \Phi \left(
    \mbox{\boldmath $x$}\right), h_{ij} \left( \mbox{\boldmath
      $x$}\right),\tau\right],
\label{sep}
\en and solve the zeroth order using a Bohmian approach
\cite{Holland93,Holland95}, where actual trajectories can be
calculated.  In the case of a perfect fluid, the Bohmian quantum
trajectory for the scale factor reads \cite{PPP07}
\begin{equation}
  \label{at} a(\tau) = a_0
  \left[1+\left(\frac{\tau}{T_0}\right)^2\right]^{1/[3(1-\omega)]} ,
\end{equation}
where $a_0$, the value of the scale factor at the bounce\footnote{The
  background wave function at the bounce, which is a Gaussian centered
  at the singular point $a=0$, gives the probability of having a
  particular value for $a_0$, and it is very low when evaluated at
  sufficient big values of $a_0$ that can describe the large Universe
  in which we live.  However, if one takes background wavefunctions at the
  bounce consisted of Gaussians traveling away from the singular point
  $a=0$ with speed parameter $u$, this problem can be overcome and
  large Universes can be obtained with reasonable probabilities
  \cite{pintasso}.  This is an example of the fact that the scales of
  the Universe are not uniquely determined by Planck scale but also on
  parameters appearing in its quantum state}, and $T_0$ are arbitrary
constants to be eventually determined by observations, and the time
parameter $\tau$ is related to conformal time $\eta$ through
\begin{equation}
\label{jauge} \dd\eta =
\left[a(\tau)\right]^{3\omega-1} \dd \tau.
\end{equation}
Note that this solution has no singularities and tends to the
classical solution when $\tau\rightarrow\pm\infty$\footnote{As all
  quantum trajectories, and hence the mean value of $a$, have this
  same functional form, then using a probabilistic interpretation,
  like the Many Worlds interpretation \cite{Everett57}, will
  presumably give the same forthcoming results: we expect the explicit
  use of a Bohmian interpretation for quantum mechanics to be of no
  practical consequence.}.  Hence, once an initial condition has been
given, $a(\tau)$ can really be understood as a mere function of
time. This function is henceforth plugged into the Fourier mode
equations for the perturbations, where it serves as a source for
``particle production'' just as in the usual inflation
calculations. This mode equation reads \cite{PP07}
\begin{equation}
\label{equacoes-mukhanov} v''_k+\biggl(\omega
k^2-\frac{{a''}}{a}\biggr)v_k=0,
\end{equation}
where $v$ reduces to the Mukhanov-Sasaki variable \cite{MFB92} when
the background satisfies the classical Einstein equations and a prime
means derivative with respect to conformal time.  The potential
$V=a''/a$, which yields the scale of curvature of the bouncing quantum
background $\ell_{_\mathrm{C}}\equiv a V^{-1/2}$, has the same
qualitative properties as the potential for perturbations in
inflation: it is negligible when $|\eta|\rightarrow\infty$ and has its
maximum around $\eta=0$.  As an explicit example, its form for a
radiation fluid reads
\begin{equation}
  \label{potential:rad} 
  V_\mathrm{rad}=\frac{1}{T_0^2\left(1+
      \displaystyle\frac{\eta^2}{T_0^2}\right)^2}, 
\end{equation}
whereas the dust case reads
\begin{equation}
\label{potential:dust} 
V_{\mathrm{dust}}=\frac{2 a_0^2}{9 T_0^2} \displaystyle
\frac{3+x^2}{\left(1+x^2\right)^{4/3}}, 
\end{equation}
where we have set $x\equiv \tau/T_{0}$. In both cases, the potential
is vanishing in the limit $|\tau|\to\infty$, \ie far from the bounce,
and reaches its maximum at the bounce itself.  Hence, as in inflation,
scales of physical perturbations are much smaller than the curvature
scale in the far past (they are above the potential, \ie $k^2\gg V$),
where they oscillate and can be set in quantum vacuum state. When the
bounce approaches, these scales get larger with respect to the
curvature scale and eventually enter the potential ($k^2\ll V$), where
they get amplified.  Finally, they become smaller again than the
curvature scale in the far future (exit from the potential), where
they oscillate again, now amplified.  Most of the time, \ie far from
the bounce itself, the background scale factor thus obtained is
undistinguishable from the solution classical Einstein (Friedmann)
equation. As all of the effects discussed in the previous section take
place in these regimes, one can consistently assume Einstein gravity
throughout the relevant history of the Universe.

In this category of models, the index $n_{_\mathrm{S}}$ of
Eq. (\ref{indexS}) can be tuned as close to unity as one wishes, but
from above. This means the spectrum is expected to be slightly blue,
as opposed to at least the simplest single field inflationary models
in which it is slightly red. The latest WMAP3 \cite{WMAP3,Komatsu+08}
observations do not currently favor this prediction but do not rule
it out either, especially if $n_{_\mathrm{S}}$ is sufficiently close
to 1 \cite{KKMR06}.
    
The amplitude of the perturbations needs be calculated
numerically. The free parameters of the background must then be
adjusted in order to fit observational data and theoretical
consistency and completeness constraints.  On the observational side,
one must have a background compatible with the large Universe we see
today and perturbations which fit the CMB data. As for the
theoretical issues, one must impose that the gauge invariant variables
always remain in the linear regime and, at least in principle, that
scales of cosmological interest were never smaller than the Planck
length in order to avoid any trans-Planckian problem by construction.

Taking into account the constraints on the parameters due to the
normalization conditions and the compromise that the model should
describe our real Universe in fact leads to imposing that the scale
factor at the bounce must be large in Planck units.  Once this is
done, there is no trans-Planckian problem \cite{MB01,BM01} and no
departure from linearity.

Finally, we should like to emphasize a major, possibly observable in
the future, difference between inflation and such bouncing models: the
so-called consistency \cite{MFB92} relation between the
tensor-to-scalar ratio $T/S$ and the spectral index. While a typical
inflation predication is a linear relation, the bounce case, on the
other hand, predicts \cite{PPP07} $T/S \propto
\sqrt{n_{_\mathrm{S}}-1}$. In the case the scalar index is very close
to 1, which is the current best fit with WMAP data \cite{KKMR06},
then $T/S$ would be very small.  Further improved data, notably on
$B$modes in the CMB, will provide a very stringent, and hopefully
discriminating, test as they will have the ability to provide a
measure of $T/S$ up to values \cite{ACK07} of order $10^{-3}$.

\section{Conclusions}

The theory of linear quantum perturbations has been successfully
applied in the framework of a classical inflationary background: only
the perturbations were quantized, leading to a sort of semiclassical
approximation to quantum gravity \cite{MC812}. We have developed a
Hamiltonian formalism where not only the perturbations but also the
background could be quantized \cite{PPP05,PPP06,PP07,PPP07}.  This led
to a picture of quantum perturbations evolving in a nonsingular
bouncing background spacetime from a vacuum state yielding spectral
indices and amplitudes that can be made to agree with observations
provided the dominant fluid in the background when the perturbation
scale becomes smaller than the curvature radius is dust. The curvature
scale at the bounce can always be set larger than the Planck length,
and hence the calculations are not spoiled by higher order quantum
gravity effects. Finally, such a model can be extended to include a
radiation-dominated decelerating phase before nucleosynthesis without
corrupting its main features properties.  This thus provides a simple
theoretical framework where only the basic principles of general
relativity and quantum mechanics, together with the assumption of the
existence of a dustlike fluid (dark matter?), yield what can be
argued to be a sensible model. Furthermore, such behaviors can also be
obtained in other nonquantum bouncing models \cite{FPP,BV05,FB02},
indicating that these are not particular properties of the specific
models here discussed.

We have also argued that general bounces provide different
perspectives on old issues such as flatness and homogeneity. In fact,
these problems may be alleviated or solved using simple physical
arguments which can be applied only in this context.

There are, however, many open questions left to be addressed and some
weak points. Let us list them below.

$\to$ Baryogenesis and dark energy are not addressed, but the latter
could actually provide a means of obtaining a redder spectrum.

$\to$ Was radiation always there, or it was produced at the bounce, \eg
through the evaporation of mini black holes or monopolonium bound
states?  Does its presence alter the amplitude of the perturbations,
and if so, how?

$\to$ As primordial perturbations are enhanced at the bounce,
similarly one could think that they also might lead to large amounts of
particle production.  The relic density of these particles needs
be evaluated for each model.

$\to$ Although spatial curvature is expected to be negligible during
most of the evolution, particularly in the expanding phase, it may be
quite important at the bounce itself and modify the amplitude of the
perturbations.

All of the properties of bouncing models and their open issues show that
they seem to provide a robust alternative to inflation. A less
ambitious role, although still very important, should be that they can
complement inflation by solving the singularity problem, ease the
homogeneity problem and yield appropriate initial conditions for it
\cite{FLP08}. In any case, bounce cosmology leads to numerous new,
hopefully measurable \cite{LPRprep}, ideas and effects, yet to be
investigated. The tensor-to-scalar prediction is already an example of
such an effect rendering the paradigm testable.

As a final remark, we would like to stress that, in contradistinction
with models in which time begins, there is no point to asking what
the probability is of the appearance of some particular eternal model out of
nothing. Contrary to the usual perspectives, one can as well assume
existence to be conceptually prior to nonexistence, \ie existence
itself may not be deserving explanation.  This is the idea underlying
our category of models: the Universe always existed and its
``appearance'' is thus a non question.

\section*{Acknowledgements} We would like to thank CNPq of Brazil for
financial support. We would also like to thank both the Institut
d'Astrophysique de Paris and the Centro Brasileiro de Pesquisas
F\'{\i}sicas, where this work was done, for warm hospitality. We very
gratefully acknowledge various enlightening conversations with Martin
Lemoine, Andrei Linde, J\'er\^ome Martin and Slava Mukhanov. We also
would like to thank CAPES (Brazil) and COFECUB (France) for partial
financial support.

\end{document}